\documentclass[conference]{IEEEtran}

\ifCLASSINFOpdf
\else
\fi
\usepackage{setspace}
\usepackage{graphicx}
\usepackage{cite}
\usepackage{amsmath}

\begin{document}

\title{On the Performance of Low Density Parity Check Codes for Gaussian Interference Channels}

\author{\IEEEauthorblockN{Shahrouz Sharifi}\\
\vspace*{-.1in}
\IEEEauthorblockA{School of  Electrical, Computer and Energy Engineering\\ Fulton Schools of Engineering\\
Arizona State University, Tempe, AZ 85287--5706, USA\\
Email: sh.sharifi@asu.edu\vspace*{-.38in}}   \and \IEEEauthorblockN{Tolga M. Duman}\\
\vspace*{-.1in}
\IEEEauthorblockA{Department of Electrical and Electronics Engineering\\ Bilkent University, Bilkent, Ankara, 06800, Turkey\\
Email: duman@ee.bilkent.edu.tr}
}

\maketitle

\begin{abstract}\let\thefootnote\relax\footnotetext{This research is primarily funded by the National Science Foundation under the contract NSF-CIF 1117174. T. M. Duman's work is also supported by the European FP7 NEWCOM$\#$ Program.}
\let\thefootnote\relax\footnotetext{T. M. Duman is currently with Bilkent University in Turkey, and on leave from Arizona State University, Tempe, AZ.}
In this paper, two-user Gaussian interference channel~(GIC) is revisited with the objective of developing implementable (explicit) channel codes. Specifically, low density parity check~(LDPC) codes are adopted for use over these channels, and their benefits are studied. Different scenarios on the level of interference are considered. In particular, for strong interference channel examples with binary phase shift keying~(BPSK), it is demonstrated that rates better than those offered by single user codes with time sharing are achievable. Promising results are also observed with quadrature-shift-keying~(QPSK). Under general interference a Han-Kobayashi coding based scheme is employed splitting the information into public and private parts, and utilizing appropriate iterative decoders at the receivers. Using QPSK modulation at the two transmitters, it is shown that rate points higher than those achievable by time sharing are obtained. 
\end{abstract}

\IEEEpeerreviewmaketitle

\section{Introduction}

There is a large body of work on Gaussian interference channels~(GIC) from an information theoretic viewpoint. Capacity characterizations of GIC have been pursued for more than three decades; however, capacity region is still unknown in general. For some special cases, for instance, under strong interference, interference channel capacity region is characterized~\cite{sato1981capacity}.  The best known achievable rate region for two-user GIC is due to Han and Kobayashi~\cite{Kobayashi1981} where information of each user is split into two parts, i.e. private and public messages. Private messages are decodable at the intended receiver while public messages can be decoded at both receivers. This intellectual partitioning gives the flexibility to achieve different rate points by allocating the public and private message powers depending on the level of interference. For instance, when the interference power is higher, more power should be allocated to the public messages. Some other recent results on rate regions and outer-bounds for GIC have been reported in~\cite{etkin2008gaussian,Annapureddy2009}.


For different multi-user channels, such as multiple access, relay and broadcast channels, in addition to the information theoretic advancements, specific channel coding techniques (explicit and implementable) have been studied. Thanks to their excellent performance, capacity-approaching LDPC codes have been designed for two users multiple access channels using Gaussian approximations and EXIT chart studies~\cite{Roumy2007,Balatsoukas-Stimming}. Codes approaching information theoretical limits have been reported for Gaussian broadcast channels and for slow fading scenarios 
based on dirty paper coding~\cite{Uppal2009}. Relay channels have also been investigated and optimized codes have been obtained with the aid of EXIT chart analysis~\cite{Hu2007}. Also for the relay channels, bi-layer LDPC codes performing well at two different SNRs are developed~\cite{Razaghi2006}. Although there are explicit channel coding results for different multi-user channels, in the existing literature, there are no results on practical (explicit) coding schemes for the interference channels which is the main motivation of this paper.  

In this paper, two-user Gaussian interference channels with fixed channel gains are considered. Since continuous and Gaussian distributed signaling is not feasible, constrained constellations, such as phase-shift-keying~(PSK) are utilized. The main idea is to implement in a practical manner the so-called ``Han-Kobayashi'' encoding procedure using LDPC codes. While the coding approach is described for the general case, two different scenarios on the level of interference are considered to provide specific examples: strong interference and general interference. Under strong interference, messages of both users should be public since both messages can be decoded at both receivers. Thus, two-user GIC under strong interference can be treated as two multiple-access-channel~(MAC)s. Consequently, optimized codes designed for MAC~\cite{Roumy2007} can be utilized when interference power is strong. However, new codes should be designed when the interference is not strong enough, e.g. under general interference. In this case, portion of the power should be allocated to the private messages because all of the interference can not be decoded and cancelled out. 

In order to illustrate the advantages of proposed explicit channel coding techniques, two LDPC codes designed for (point to point) additive white Gaussian noise~(AWGN) channels with different rates are employed~\cite{Brink2004,website:mackay}. Three different examples are studied in detail. In the first scenario simulation results suggest that by using simple BPSK constellations, one can do better than single user codes used with time sharing. In the second scenario, QPSK is utilized and it is shown that higher rates compared to single user codes (with Gaussian alphabets) with time sharing can be achieved. In the third (general interference) scenario it is observed that by using an appropriate power allocation to public and private messages at the transmitter side, QPSK signaling is capable of achieving higher rates than single user codes with time sharing. These results clearly demonstrate the potential of LDPC codes for use over interference channels. Also noting that these encouraging results are attained with off-the shelf codes adopted in a Han-Kobayashi encoding scheme, even without optimizing the LDPC codes for the interference channels. This motivates us to further investigate design of channel codes which can operate closer to the known achievable rate region boundaries of the interference channel. 

The rest of the paper is organized as follows. In Section II, Gaussian interference channel model is described. In Section III, coding and decoding schemes adopted from Han-Kobayashi type coding are explained in detail. In Section IV, several numerical examples are given which illustrate the potential of the proposed LDPC coding solutions. Finally, in Section V, conclusions are provided.

\section{System Model} 

\begin{figure}[h]
\centering
\includegraphics[scale=.45]{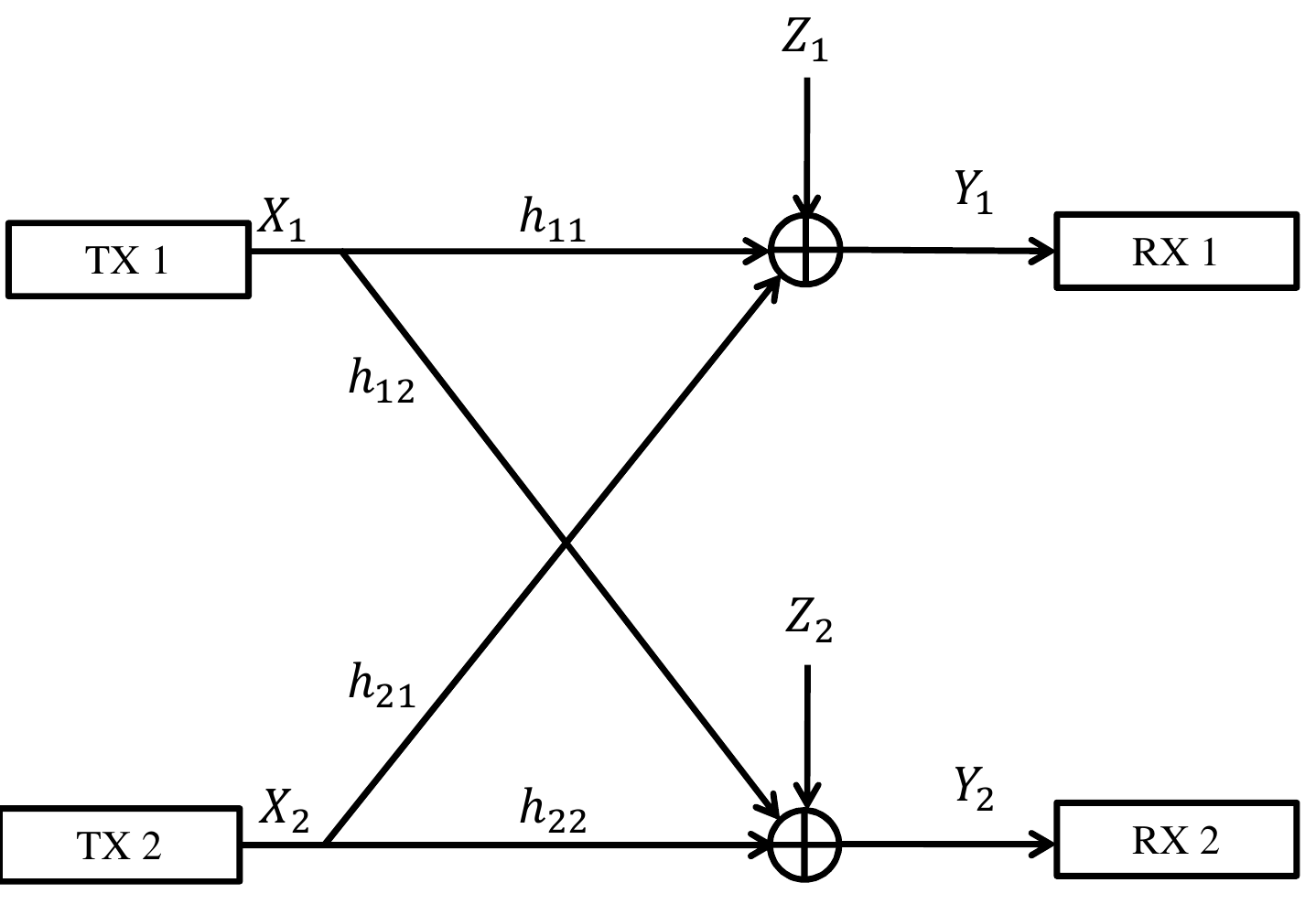}
\caption{Interference Channel~(IC)}
\label{IC}
\end{figure}

Two-user GIC system model is shown in Fig.~\ref{IC}. Channel outputs can be written as
\begin{equation}
\begin{array}{l}
Y_1=h_{11}X_1+h_{21}X_2+Z_1,\\
Y_2=h_{12}X_1+h_{22}X_2+Z_2,
\label{GICmodel}
\end{array}
\end{equation}
where $h_{ij}$ is the channel gain from user $i$ to receiver $j$ and $Z_1$ and $Z_2$ are independent and identically distributed ~(i.i.d) noise processes~($\mathcal{CN}(0,N_0)$). $X_1$ and $X_2$ are transmitted signals with power constraints of $P_1$ and $P_2$, respectively. That is, $E\{|X_i|^2\}=P_i$~($i=1,2$). Signal-to-noise-ratio~(SNR)s and interference-to-noise-ratio~(INR)s at receiver $i$ are defined as
\begin{equation}
\mathrm{SNR}_i=\frac{|h_{ii}|^2P_i}{N_0},~\mathrm{INR}_i=\frac{|h_{ji}|^2P_j}{N_0},~i,j=1,2~~i\neq j.
\end{equation}
In the symmetric interference channel we have
\begin{equation}
\begin{array}{c}
|h_{11}|=|h_{22}|,~~|h_{12}|=|h_{21}|\\
\mathrm{SNR}_1=\mathrm{SNR}_2=\mathrm{SNR},\\
\mathrm{INR}_1=\mathrm{INR}_2=\mathrm{INR},
\end{array}
\end{equation}
where strong interference correspond to $INR>SNR$.

\section{Coding and Decoding Schemes} 

\subsection{Encoding}

Considering Han-Kobayashi coding scheme, message of each user is divided into two parts, namely, private~($U$) and public messages~($W$). Public messages can be decoded at both receivers while the private messages are only decodable at the intended receivers. Although in the general scheme information data is split into public and private messages, there are special cases where there is no need to allocate power to private messages. For instance, under strong interference, messages of both users can be decoded at the receivers. This gives the intuition that both users use all the power for public messages which makes the channel look like as two MACs.

\begin{figure}[h]
\centering
\includegraphics[scale=.5]{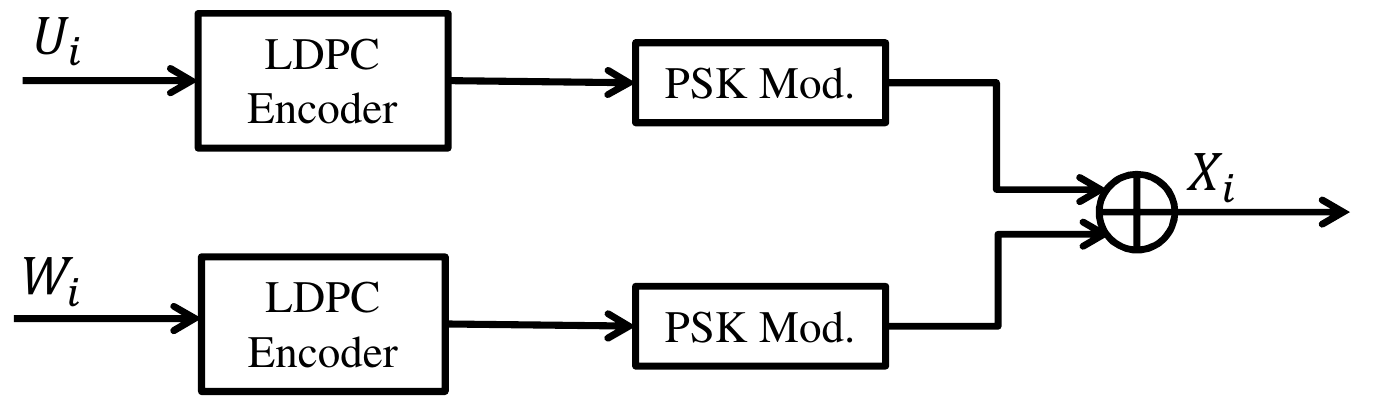}
\caption{Transmitter side block diagram~(Han-Kobayashi coding scheme)}
\label{HKscheme}
\end{figure}

Fig.~\ref{HKscheme} shows block diagram of the transmitter incorporating Han-Kobayashi coding scheme. As shown in the figure, message of each transmitter is split into public and private parts, and each are encoded using LDPC codes. The encoded sequences are then modulated using PSK, and superimposed to form the overall signal to be transmitted. In this paper we superimpose the two signals with standard addition; however, it is also possible to consider other possibilities. For instance, binary addition of the two LDPC coded words can be used superimposing the two signals for the public and private messages in the ``code'' domain. As another example, it is also possible to consider higher order signal constellations, and perform mappings of the public and private coded bits to the constellation points jointly. We note that the focus is on practical signal constellations such as PSK signals since using Gaussian signaling (as usually done in information theoretic studies) results in schemes very difficult to decode in an implementable manner. 

\subsection{Decoding}

\begin{figure}
\centering
\includegraphics[scale=.5]{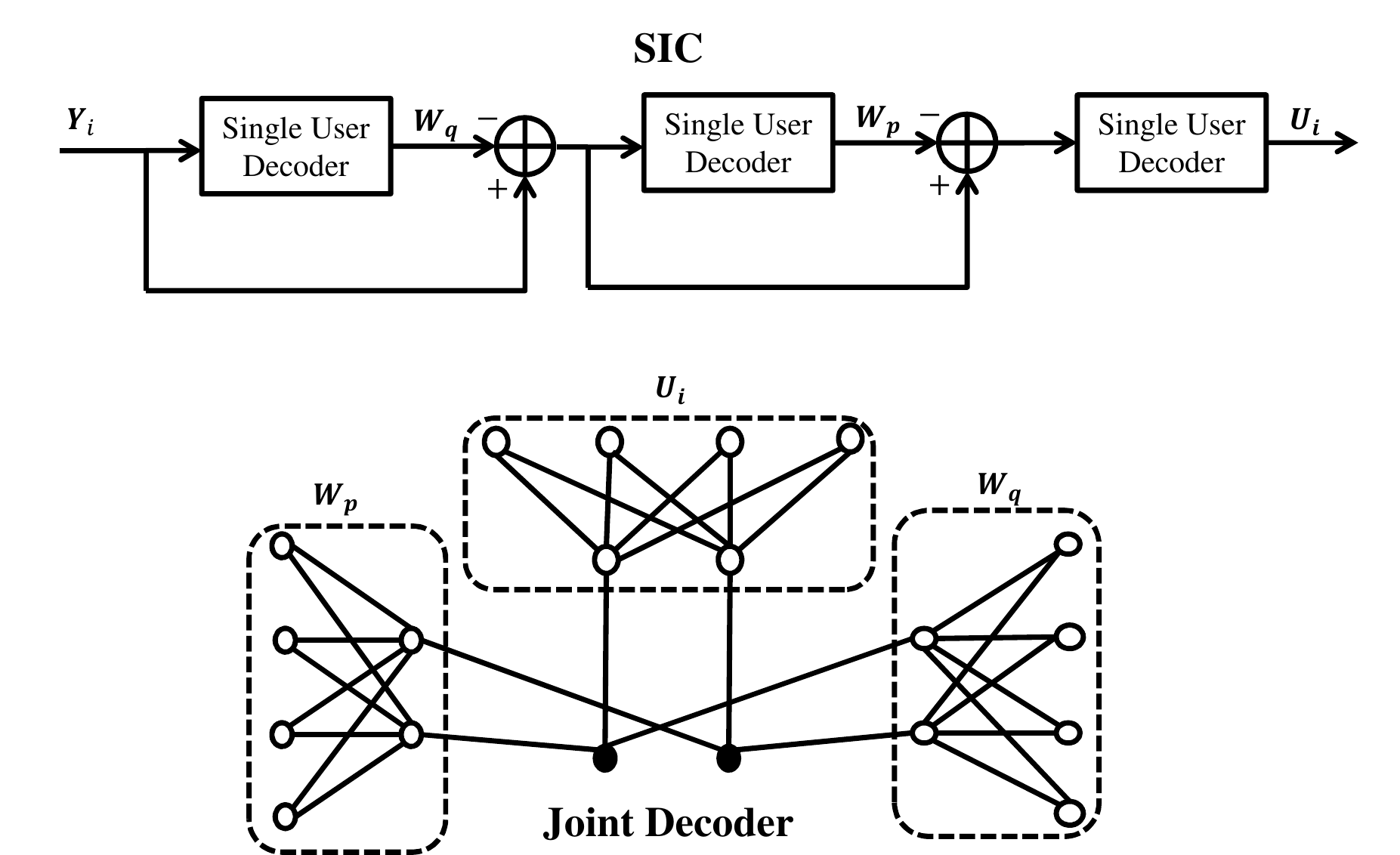}
\caption{SIC and joint decoder block diagrams~($i,p,q=1,2,~p\neq q$)}
\label{SIC-joint}
\end{figure}

At the receiver side, public messages and the private message of the desired user can be decoded jointly or successively~(illustrated in~Fig.~\ref{SIC-joint}) employing conventional message passing algorithm called Belief-propagation~(BP) algorithm~\cite{Richardson2001c}. In joint decoding, decoding of the messages are performed simultaneously using a joint decoding graph where soft messages are exchanged utilizing parallel or serial message passing scheduling (similar to~\cite{Roumy2007} for MAC). In parallel scheduling, decoding can be done for all sub-graphs corresponding to different messages whereas in serial scheduling decoding is performed only for one of the sub-graphs at each iteration. At the end of each iteration, soft information is exchanged through the intermediate nodes called state nodes improving decoding by running multiple iterations.

In successive interference cancellation~(SIC), unlike joint decoding, decoding is done sequentially using single user decoders. Decoding process initiates with the strongest message which can be a public or a private message depending on the level of the interference and power allocations at the transmitter sides. Decoded messages are then subtracted out from the original signal and fed to the next single user decoder until all messages are decoded. It is also possible to iterate between these decoders by passing hard decision information obtained from previous decoded messages to improve the decoding performance. This approach can be thought as a specific instance of joint decoding; however, we differentiate them by the use of hard or soft information being passed between among different modules. Since soft information is being passed, the joint decoding in general has a better performance while SIC is simpler to implement. Here we adopt the SIC approach due to its simplicity.

Considering BP, log-likelihood-ratio~(LLR)s of the $i$th coded bit of user $j$, denoted as $c^j_i$, at receiver one fed to the single user decoder is computed as follows:
\begin{align}
L(c^j_i)&=\log\left(\frac{P(c^j_i=0|y_1)}{P(c^j_i=1|y_1)}\right) \nonumber \\
 & = \log\left(\frac{P(y_1|x^j_i=1)}{P(y_1|x^j_i=-1)}\right) \nonumber \\
&=\log\left(\frac{\sum_{X^j\in S^{j+}_i}^{}P(y|X^j)P(X^j)}{\sum_{X^j\in S^{j-}_i}^{}P(y|X^j)P(X^j)}\right),
\end{align}
where $x^j_i$ and $X^j$ are the $i$th coded bit and user $j$'s codeword, respectively. $S^{j+}_i$ and $S^{\prime j-}_i$ are the set of all codewords of user $j$'s bits over a block assuming $x^j_i=1$ and $x^j_i=-1$, respectively. 

In SIC, there is no knowledge of the messages that have not been decoded yet. As a result, $P(X^j)$ is constant throughout decoding of each message. However, in joint decoding $P(X^j)$ is updated through the decoding with the knowledge of other user's LLRs passed to the state nodes. 

\subsection{Achievable Rate Regions}

\begin{figure}[h]
\centering
\includegraphics[width=7.2cm, height=7.2cm]{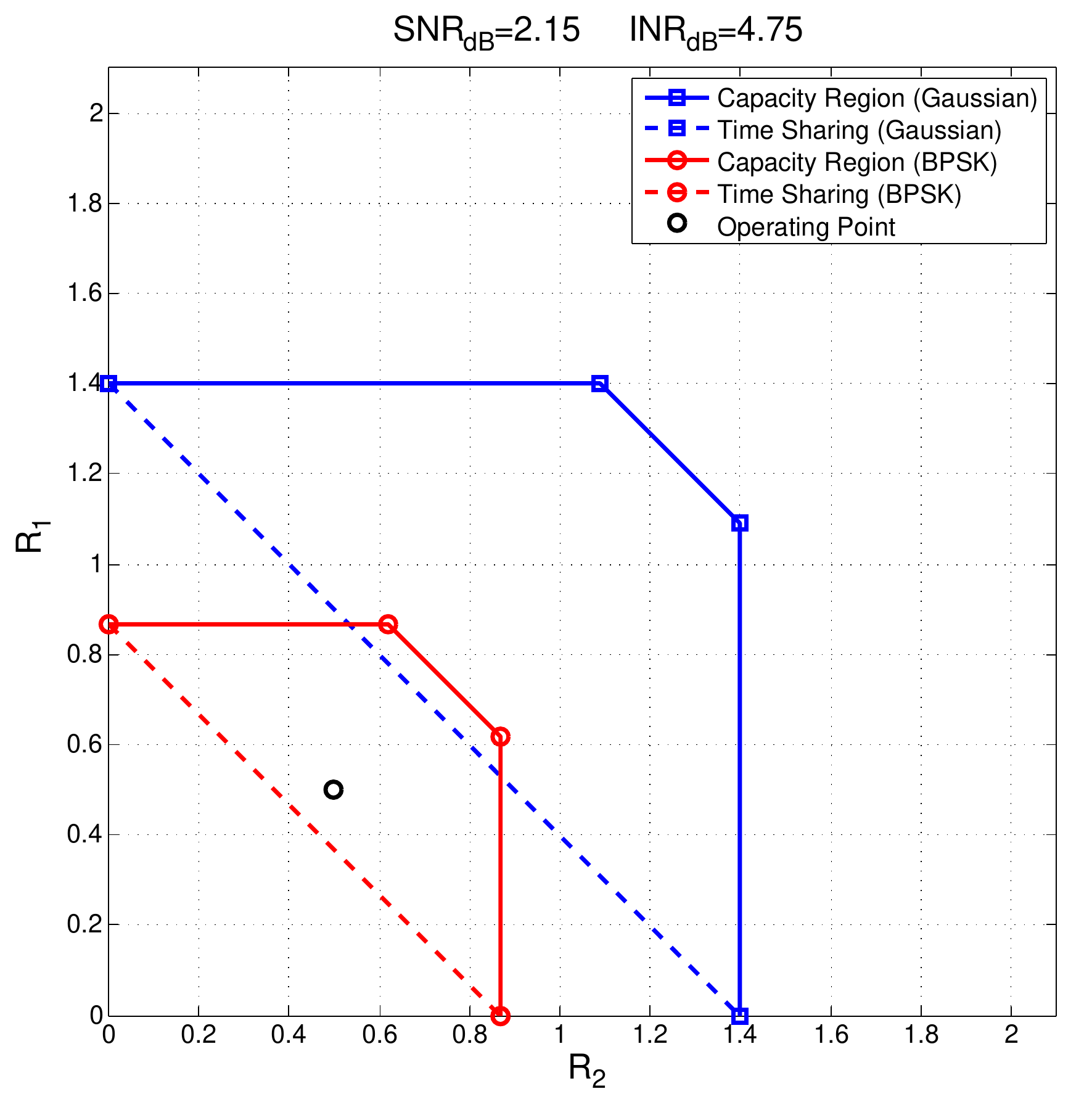}
\caption{Scenario I:~Gaussian and BPSK capacity regions~(Operating point $(R_1,R_2)=(0.5,0.5)$).}
\label{5-bpsk}
\end{figure}

\begin{figure}[h]
\centering
\includegraphics[width=7.2cm, height=7.2cm]{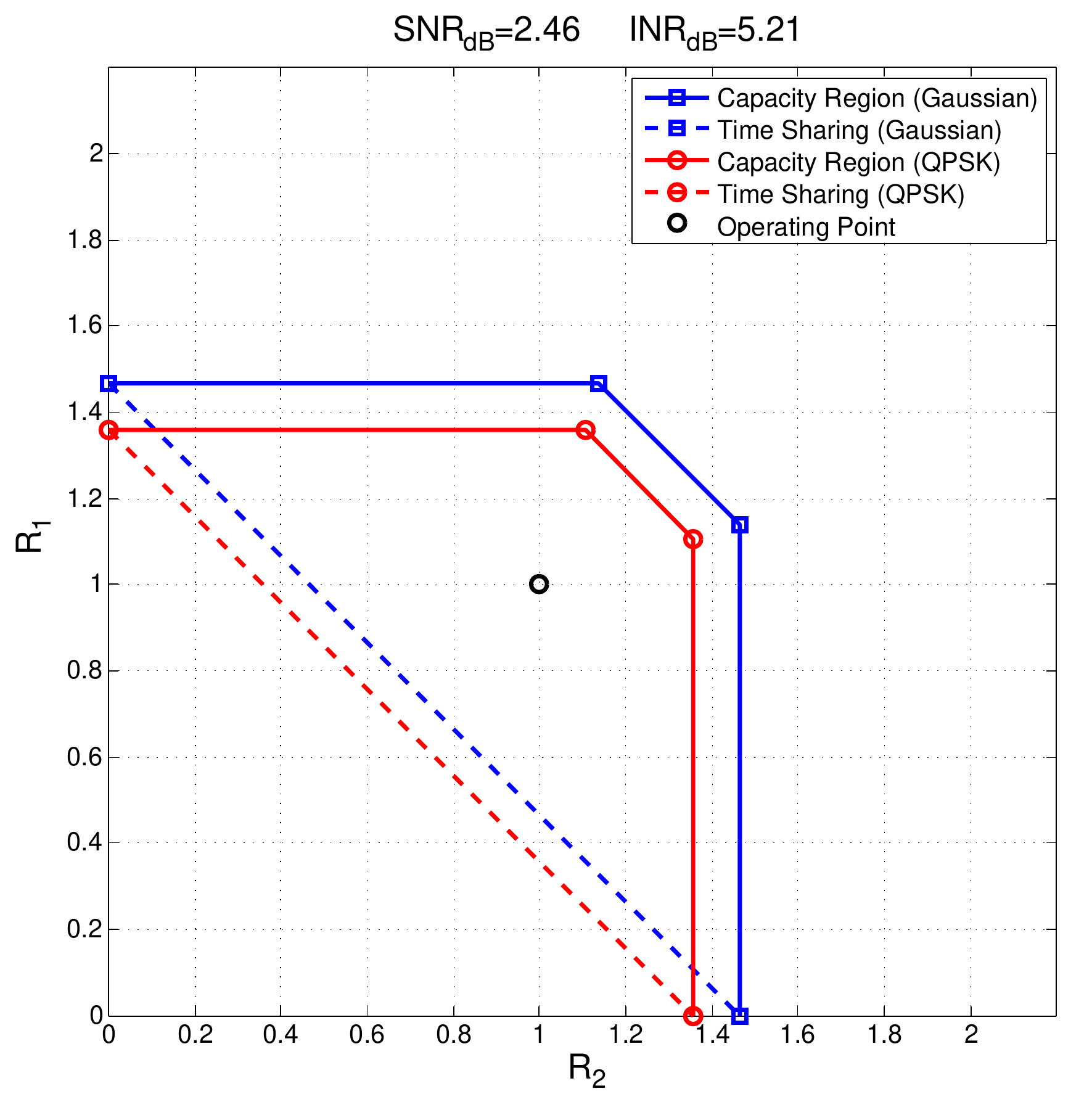}
\caption{Scenario II:~Gaussian and QPSK capacity regions~(Operating point $(R_1,R_2)=(1,1)$).}
\label{5-qpsk}
\end{figure}

Capacity region is not known for interference channels in general and only achievable rate regions and outer bounds are available. For the special case of a strong interference channel, all the information is transmitted through public messages, and both users are able to decode each others' messages. The strong interference channel can be viewed as two MAC channels and the capacity region is characterized for both Gaussian and finite constellations~\cite{sato1981capacity,Harshan2011}. Fig.~\ref{5-bpsk} and~\ref{5-qpsk} display capacity regions computed for BPSK and QPSK alphabets using Monte Carlo type calculations which match with the formulations given in~\cite{Harshan2011}. 

\begin{figure}[h]
\centering
\includegraphics[width=7.2cm, height=7.2cm]{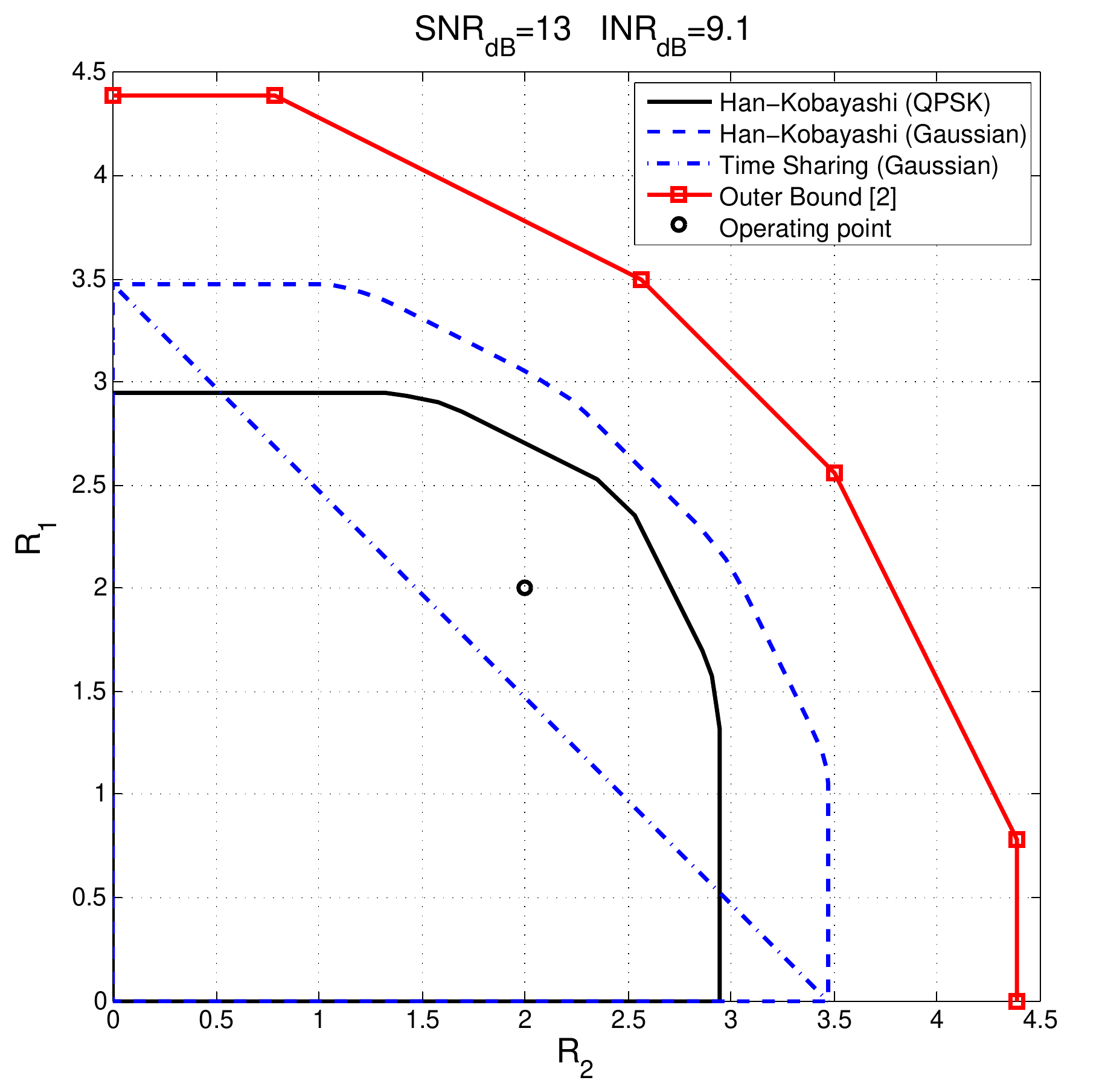}
\caption{Achievable rate regions and the outer bound.}
\label{convex}
\end{figure}

As previously mentioned, an achievable rate region over a general interference channels is developed in~\cite{Kobayashi1981} named the Han-Kobayashi rate region; however, computation of the entire region is difficult since one should perform optimization over probability-mass-function~(PMF)s of many random variables with large cardinalities. Authors in~\cite{Chong2008} have proposed a simplified version of the rate region which still requires many computations. In this paper, the focus is on Gaussian interference channels, and instead of computing the entire region~\cite{Kobayashi1981}, a subregion is computed with a smaller complexity by assuming that public and private messages are encoded using codebooks with independent and identically distributed elements. Since different power allocations to public and private messages give rise to different sub-regions, convex hull of all the sub-regions are considered (e.g. through the use of time-sharing). As a specific example, Fig.~\ref{convex} demonstrates the convex hull of a sub-region computed for power allocations ranging from $0$ to $10$, i.e. $\frac{P_u}{P_w}=\{0,...,10\}$ using Monte Carlo simulation. Gaussian achievable rate region along with the time sharing line is plotted using the formulation given in~\cite{Kobayashi1981}. Since the actual capacity region is unknown, the outer bound provided in~\cite{etkin2008gaussian} is also shown in the figure.

\section{EXAMPLES ON LDPC CODE PERFORMANCE OVER GIC}

We now provide several examples of LDPC code performance for different Gaussian interference channels. To simplify the exposition, we use symmetric Gaussian interference channels with constant channel gains. Users encode their information bits with identical LDPC codes. Two irregular LDPC codes optimized for point to point AWGN channels~\cite{Brink2004,website:mackay} are exploited with rates $0.25$~(block length~13298) and $0.5$~(block length~10000), respectively. Encoded messages are modulated using BPSK and QPSK constellations. SIC is adopted at the receivers. Due to the symmetry, decoding results are provided for only one of the receivers with the understanding that the other one enjoys the same performance. Three different scenarios are considered. In Scenarios I and II, BPSK and QPSK are used over two strong interference channels. In Scenario III, QPSK is used over a general interference channel. All simulations are performed for $10M$ information bits. A bit error probability of $10^{-4}$ is considered as reliable transmission, hence we say that the ``operating rate pair'' is achieved. 

\subsubsection{Scenario I~--~BPSK over a Strong Interference Channel}

\begin{figure}[t]
\centering
\includegraphics[width=7.2cm, height=7.2cm]{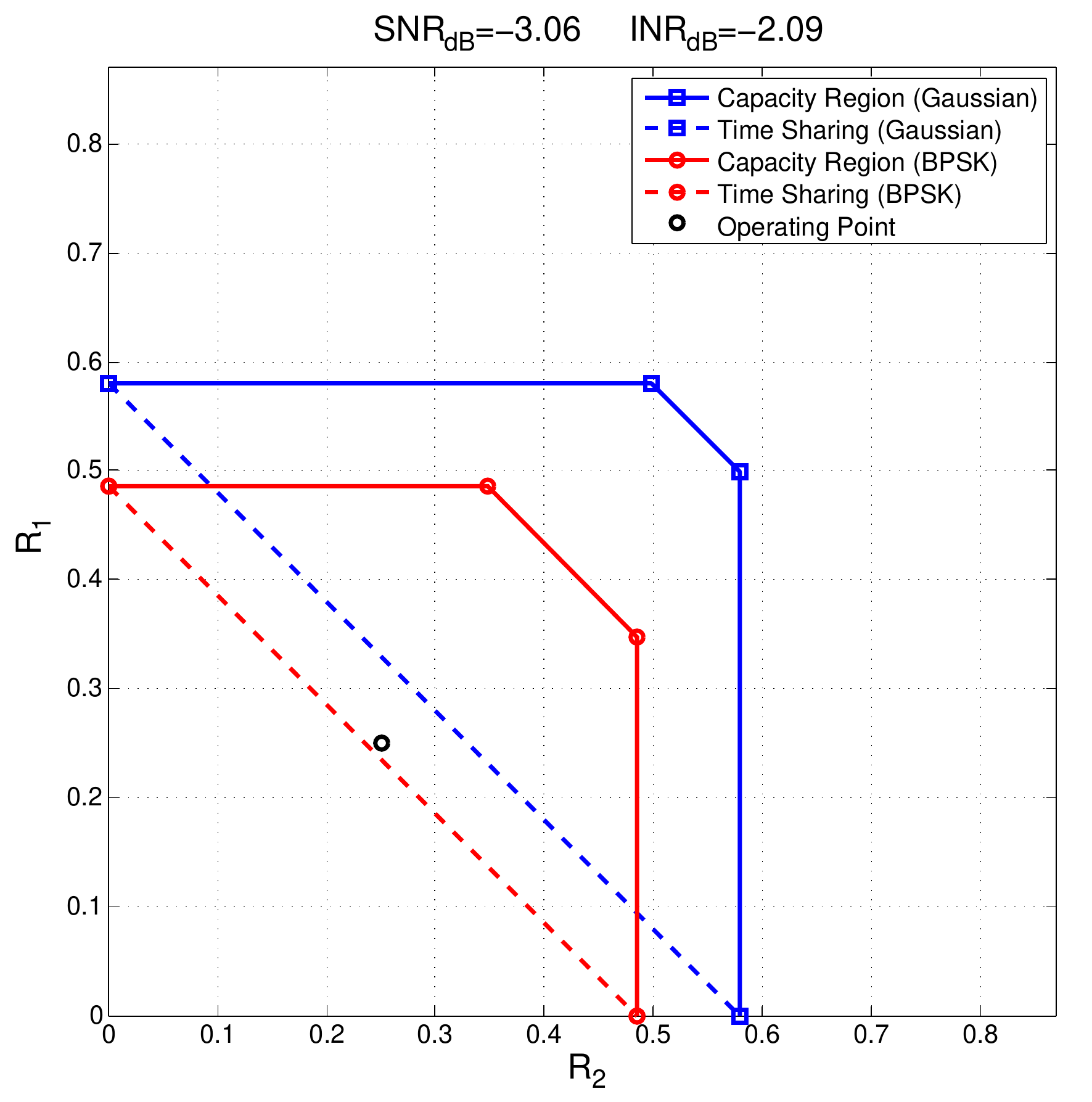}
\caption{Scenario I:~Gaussian and BPSK capacity regions~(Operating point $(R_1,R_2)=(0.25,0.25)$).}
\label{25-bpsk}
\end{figure}

In this example BPSK is used over a strong interference channel. Figs.~\ref{5-bpsk} and~\ref{25-bpsk} show the capacity regions for different SNR and INR values. The channel parameters and simulation results are shown in Table~\ref{tab:bpsk}. Considering BPSK time sharing line and specific operating points in the figures, it is clear that rate pairs achieved cannot be obtained with single user codes along with time sharing (even with infinite block length codes), hence the benefit of the specific coding/decoding schemes adopted in this paper for the interference channel scenario is clear.

\begin{table}[h]
  \centering
  \caption{Decoding Results~(Scenario I)}
    \begin{tabular}{|r|c|c|c|}
    \hline
         Code Rate & h$_{11}$, h$_{21}$ & BER$_{2}$ & BER$_{1}$  \\ \hline
    \multicolumn{1}{|c|}{$0.25$} & $0.35e^{j\frac{\pi}{4}}$, $0.44e^{j\frac{\pi}{4}}$ & $0.00013$ & $0.00012$ \\ \hline
    \multicolumn{1}{|c|}{$0.5$} & $1.16e^{j\frac{\pi}{4}}$, $2.11e^{j\frac{\pi}{4}}$ & $0.000007$ & $0.000007$  \\ \hline
    \end{tabular}%
  \label{tab:bpsk}%
\end{table}%

\subsubsection{Scenario II~--~QPSK over a Strong Interference Channel}

\begin{figure}[t]
\centering
\includegraphics[width=7.2cm, height=7.2cm]{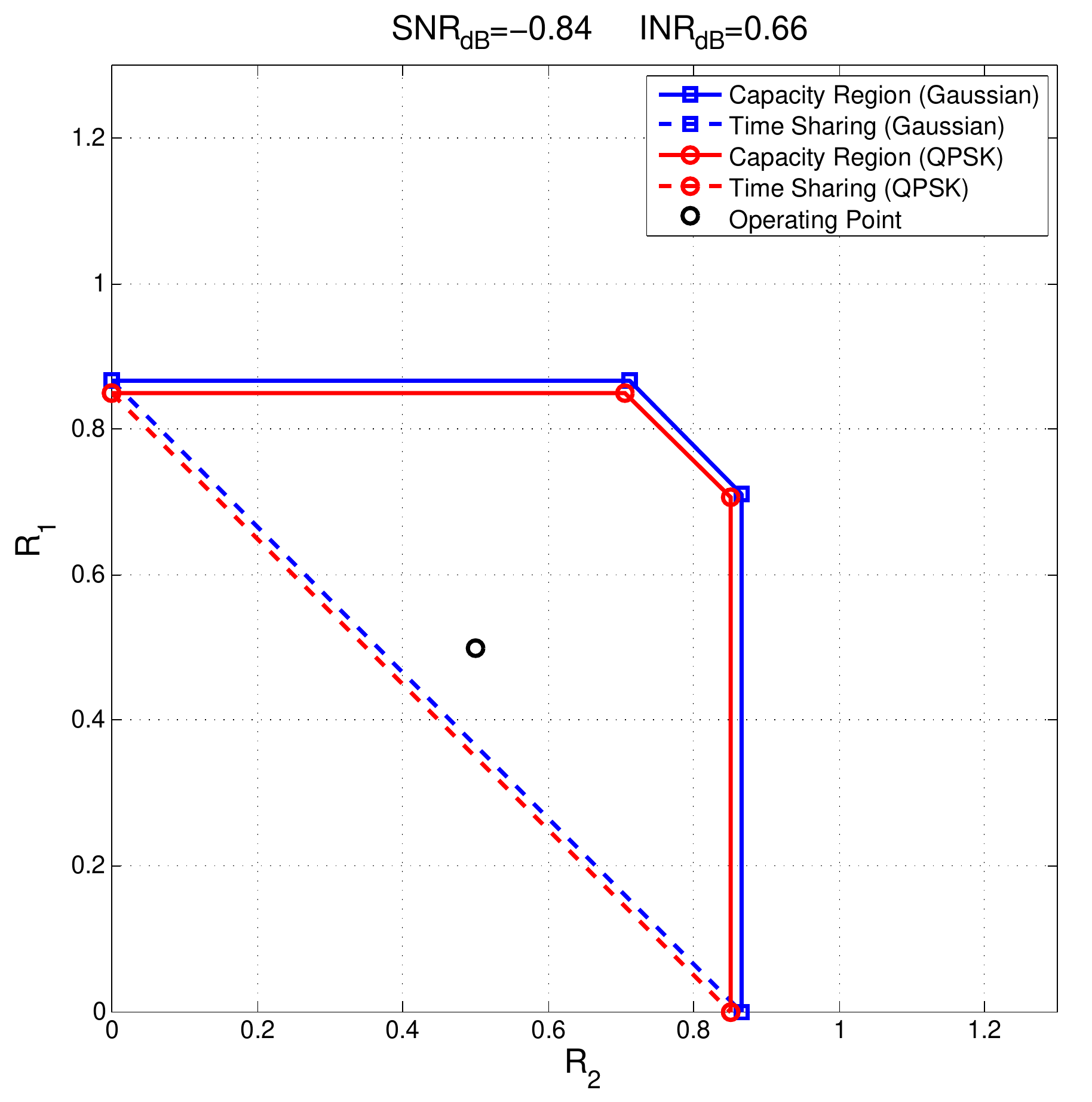}
\caption{Scenario II:~Gaussian and QPSK capacity regions~(Operating point $(R_1,R_2)=(0.5,0.5)$).}
\label{25-qpsk}
\end{figure}

In this scenario QPSK signaling over a strong interference channel is considered. Operating points and capacity regions are shown in Figs.~\ref{5-qpsk} and~\ref{25-qpsk} for different SNR and INR values. Details of the channel parameters and simulation results are shown in Table~\ref{tab:qpsk}. As the decoding results in the table indicate, the bit error rate for operating points above the time sharing line are considerably low. This shows that the offered performance is better than those of (optimal) single user codes with time sharing even with the use of Gaussian alphabets. 

\begin{table}[h]
  \centering
  \caption{Decoding Results~(Scenario II)}
    \begin{tabular}{|r|c|c|c|c|}
    \hline
         Code Rate & h$_{11}$, h$_{21}$ & BER$_{2}$ & BER$_{1}$  \\ \hline
    \multicolumn{1}{|c|}{$0.25$} &  $0.58e^{j\frac{\pi}{4}}$, $0.82e^{j\frac{\pi}{4}}$ & $0.00014$ &  $0.00014$  \\ \hline
    \multicolumn{1}{|c|}{$0.5$} & $1.25e^{j\frac{\pi}{4}}$, $2.35e^{j\frac{\pi}{4}}$ & $0.00019$ &  $0.00029$  \\ \hline
    \end{tabular}%
  \label{tab:qpsk}%
\end{table}%

\subsubsection{Scenario III -- QPSK over a General Interference Channel}

In this scenario QPSK is considered for a general interference channel example where INR is less that SNR. All  messages (private and public) are encoded employing identical LDPC codes~($N=10000, R=0.5$) with the distribution given in~\cite{Brink2004}. Unlike previous scenarios, some of the power is allocated to private messages since all the interference can not be decoded and canceled out at the receivers. Having simulated multiple power allocations and channel parameters for the SNR and INR in Fig.~\ref{convex}, following parameters resulted in considerably low bit error rates 
\[h_{11}=14.11e^{j\frac{\pi}{4}},~~h_{21}=5.75e^{j\frac{\pi}{4}},~~\frac{P_{U}}{P_{W}}=0.16,\] for this particular operating point.
Decoding results are displayed in Table~\ref{tab2} and suggest that rate pair $(2,2)$, which is shown in Fig.~\ref{convex}, is achieved. This indicates that one can do better than single user codes~(even with Gaussian alphabets) used with time sharing, and approaching the boundary of the Han-Kobayashi rate region computed for QPSK signaling. 

\begin{table}[h]
\centering
\caption{Decoding Results~(Scenario III)}
\begin{tabular}{|c|c|c|c|}
\hline Message & W$_1$ & W$_2$ & U$_1$  \\
\hline
  BER & $0.000009$ & $0.00002$ & $0.00004$\\ 
\hline 
\end{tabular}
\label{tab2}
\end{table}
\section{Conclusion}

In this paper, performance of the explicit LDPC codes are studied for Gaussian interference channels. The main idea is to implement the well known Han-Kobayashi type encoding using practical and implementable channel codes (as opposed to information theoretic random codes). After describing the general idea and providing some examples of rate regions, several specific examples are studied in detail in terms of bit error rates. Namely, two cases over strong interference channels (with BPSK and QPSK signaling) and one scenario with general interference (with QPSK transmission) are considered. It is demonstrated that using the proposed coding strategy, it is possible to achieve rate pairs (with explicit channel codes) not attainable even with optimal single user codes (with time sharing). Furthermore, it is possible to come close to operate ``near'' the boundary of the Han-Kobayashi rate region using practical LDPC codes. These results are obtained without optimizing the LDPC codes for the interference channel, hence we anticipate even better results with code optimization (which is the subject of further investigation).

\bibliographystyle{IEEEtran}
\bibliography{bib_new}

\end{document}